\documentclass{optica-article}

\journal{opticajournal} 

\articletype{Research Article}

\usepackage{lineno}
\usepackage{siunitx}
\usepackage{amsmath}

\begin{document}

\title{High-NA vectorial hologram optimization}

\author{Michael Wischert,\authormark{1,*} Fiona Hellstern,\authormark{1} Paul Uerlings,\authormark{1} Tilman Pfau,\authormark{1} Stephan Welte,\authormark{1} and Ralf Klemt\authormark{1}}

\address{\authormark{1}5. Physikalisches Institut and Center for Integrated Quantum Science and Technology, Universität Stuttgart, Pfaffenwaldring 57, 70569 Stuttgart, Germany\\

}

\email{\authormark{*}mwischert@pi5.physik.uni-stuttgart.de} 

\begin{abstract*} 
We present a vectorial phase-only framework for the gradient-based optimization of computer-generated holograms. Our approach addresses a critical limitation in high numerical aperture (NA) optical systems, where standard scalar approximations often experience significant fidelity degradation. We highlight the limitations of scalar diffraction models (Fraunhofer and Debye) at high NA for representative classes of target intensities, namely tweezer arrays and extended flat-top profiles. Using a differentiable Richards-Wolf forward model, we numerically demonstrate stable algorithmic convergence and the generation of highly uniform intensity profiles in the deep non-paraxial regime. At $\mathrm{NA}=0.9$, Richards-Wolf (RW) optimization reaches flat-top and tweezer uniformities of \SI{99.97}{\percent} and \SI{99.98}{\percent}, respectively, with a mean tweezer ellipticity of $\varepsilon=1.007$, while scalar forward models show target-dependent degradation in plateau fidelity or focal geometry. Furthermore, because the RW model returns the full vectorial field, the same framework can optimize objectives that depend on local polarization. We demonstrate this for a polarization-sensitive optical-dipole-potential target, where direct RW-potential optimization reduces the mean normalized residual by more than an order of magnitude when including vector and tensor light-shift terms.
\end{abstract*}

\section{Introduction} 
\label{Introduction}

Spatial light modulators (SLMs) have been widely used to control laser fields in quantum science and technology \cite{barredo2016,bluvstein2024, Knottnerus2025} and biological applications \cite{Lutz2008HolographicPhotolysis,Emiliani2005WavefrontMicroscopy} for decades. Complex optical trapping and beam shaping with SLMs rely heavily on computer-generated holograms (CGHs)~\cite{Jesacher:08}. In these settings, an SLM is typically used as a programmable pupil-plane phase element, and the CGH provides the numerical phase pattern that converts a simple incident beam into the desired focal intensity distribution.

Generating high-fidelity holograms presents two distinct challenges: selecting an accurate, numerically viable propagation method and solving the nonlinear, nonconvex inverse phase-retrieval problem. Because the phase retrieval problem is generally non-unique without sufficient additional constraints \cite{noUniqueSolution}, many numerical strategies have been developed, from iterative Fourier transform algorithms (IFTAs), such as the Gerchberg-Saxton algorithm \cite{GS_paper}, to modern gradient-based approaches including Wirtinger flow and direct gradient descent \cite{Uchiyama2024Wirtinger,CandesWirtinger2015,Liu2020}. In many CGH workflows, these algorithms are paired with Fraunhofer propagation because this forward model is simple to implement and computationally efficient. These standard workflows therefore predominantly solve the hologram generation problem in the scalar regime, even though high-NA focusing requires a vectorial description of the focal field. Tight focusing mixes the incident transverse polarization into cross-polarized and longitudinal focal-field components, which can distort diffraction-scale intensity features and produce spatially varying polarization states.
Vectorial phase-only beam shaping under tight focusing has previously been demonstrated with generalized-projection and Gerchberg-Saxton algorithms, including high-NA flat-top targets \cite{JabbourKuebler2008,JahnBokor2010}. More recently, differentiable Debye-Wolf/Richards-Wolf (RW) propagation has enabled gradient-based point-spread-function (PSF) engineering for adaptive optics \cite{VishniakouSeelig2023}.

Here, we focus on quantitative high-fidelity phase-only CGH for high-NA optical systems, as used, for example, in optical dipole traps. In such systems, model mismatch can directly affect trap uniformity, focal geometry, and polarization-dependent optical potentials. We use RW propagation both as the full vectorial evaluation model and as the differentiable forward model during phase optimization. This allows us to benchmark Fraunhofer, scalar Debye, and full RW optimization under a common RW evaluation, separating scalar high-NA pupil-weighting errors from vectorial polarization-mixing errors. We then extend intensity-only CGH to the optimization of electric-field functionals that scalar propagation models cannot evaluate self-consistently. Specifically, we use the computed vector field directly in the loss function to optimize a polarization-dependent optical dipole potential with vector and tensor light shifts. Such terms depend on the local polarization state of the focal field, so scalar propagation can provide only an intensity proxy rather than the full potential. This is relevant for high-NA neutral-atom platforms, where trap depths, differential light shifts, and motional frequencies are set by the optical potential landscape rather than by intensity alone \cite{PaperKien2012,Tomita2026AtomCamera}.

\begin{figure}[htbp]
    \centering
    \includegraphics[width=1\linewidth]{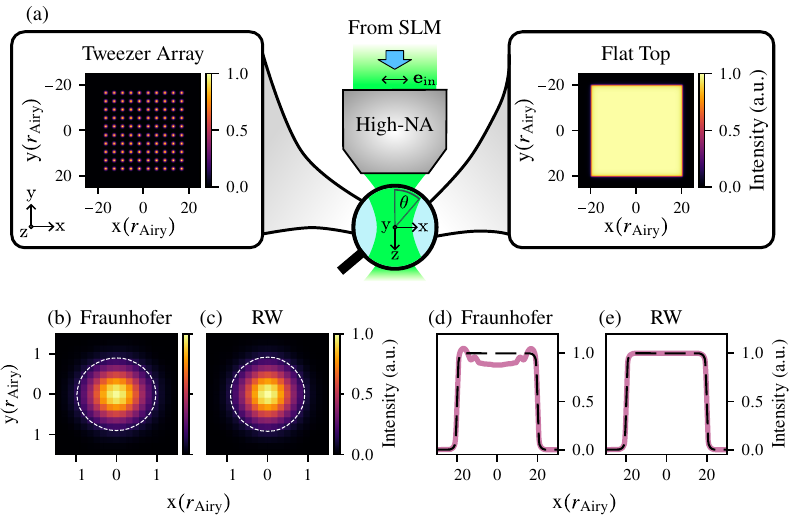}
    \caption{High-NA CGH geometry and representative Fraunhofer model errors. (a) Schematic of the SLM-based high-NA focusing geometry used for the numerical benchmarks. The insets show the tweezer-array and extended flat-top intensity targets. Here $r_{\mathrm{Airy}}=0.61\lambda/\mathrm{NA}$ denotes the Airy-radius diffraction scale. (b)-(e) All displayed intensity data are evaluated with the vectorial Richards-Wolf (RW) propagation model at $\mathrm{NA}=0.7$. (b) Magnified view of a single, near-diffraction-limited tweezer generated from a phase hologram optimized with scalar Fraunhofer propagation, showing an elliptical focal spot under RW evaluation. The dashed contour marks the $1/e^2$ intensity level. (c) Corresponding tweezer generated from a phase hologram optimized directly with RW propagation, recovering a nearly circular focal spot. (d) Line cut through the extended flat-top target for a Fraunhofer-optimized phase evaluated with RW propagation, showing strong deviation from the target profile. (e) Corresponding line cut for an RW-optimized phase, recovering the flat-top profile.}
    \label{fig:Fig1}
\end{figure}

Fig.~\ref{fig:Fig1} illustrates this scalar-model mismatch for a standard SLM-based high-NA focusing geometry and two representative target classes: a near-diffraction-limited tweezer array and an extended flat-top profile (Fig.~\ref{fig:Fig1}(a)). The deviations shown here are RW-evaluated model-mismatch errors, not failures of the scalar optimizations to converge under their own forward models. Already at $\mathrm{NA}=0.7$, phase holograms optimized with scalar Fraunhofer propagation deviate strongly from the intended targets when evaluated with the full RW model. The tweezer focus becomes elliptical (Fig.~\ref{fig:Fig1}(b)), and the flat-top profile develops strong deviations from the target uniformity (Fig.~\ref{fig:Fig1}(d)). In contrast, using the RW model directly during hologram optimization recovers a nearly circular tweezer shape and the flat-top profile (Fig.~\ref{fig:Fig1}(c) and (e)).

The remainder of the paper makes this comparison quantitative. We first formulate Fraunhofer, scalar Debye, and RW propagation in a common phase-only optimization pipeline, so that the optimized phases differ only by the forward model used to compute the loss and gradient. We then evaluate the resulting holograms with the same RW model to identify target-dependent scalar-model errors under rigorous vectorial evaluation, and we examine whether the vectorial forward model retains practical FFT-based convergence. Finally, we apply the same pipeline to a polarization-dependent optical-potential objective, using the RW field components inside the loss rather than evaluating only a scalar intensity proxy.

\section{Propagation and hologram generation} 
\label{PropagationAndCGH}
We assume that the SLM phase is imaged onto the entrance pupil of the objective and seek the phase $\varphi_{\text{SLM}}$ that creates a target intensity pattern $I_{\text{tar}}$ in the focal plane at $z=0$, as shown in Fig.~\ref{fig:Fig1}(a). The surrounding medium is assumed to have a refractive index of $n=1$ for all calculations.

\paragraph{Forward propagation models}

To accurately model high-NA focusing, we use the Richards-Wolf \cite{RWpaper} formalism, which accounts for non-paraxial diffraction and the vectorial nature of light without requiring the significant computational overhead of direct wavefield propagation over macroscopic focal distances. We adopt the coordinate conversion and pupil mapping defined by Leutenegger et al.~\cite{2006Leutenegger}, mapping the spherical reference wavefront onto planar pupil coordinates $(k_x, k_y)$. In the phase-only SLM geometry considered here, the field incident on the objective pupil is written as $\mathbf{E}_{\text{in}}(k_x, k_y) = A(k_x, k_y) \exp[i\varphi_{\text{SLM}}(k_x,k_y)]\,\mathbf{e}_{\text{in}}$, where $A$ is the fixed pupil amplitude and $\mathbf{e}_{\text{in}}$ is the transverse input Jones vector. In the simulations below, the pupil amplitude is uniform inside the circular aperture and zero outside, and the input is linearly polarized $\mathbf{e}_{\text{in}}=\hat{\mathbf{x}}$, so $E_{\text{in},y} = E_{\text{in},z}=0$. A single phase-only SLM phase therefore acts as a common scalar phase factor on the incident transverse field. The fully vectorial propagation integral in the focal region can be written as the Fourier transform ($\mathcal{F}$) of the generalized pupil function \cite{2006Leutenegger}
\begin{equation}
\label{eq:Formula_Vectorial}
\mathbf{E}_{\text{RW}}(\mathbf{r}) \propto
\mathcal{F} \left[
\frac{1}{\sqrt{\cos\theta}}
\cdot \mathbf{M}(k_x, k_y)
\cdot \mathbf{E}_{\text{in}}(k_x, k_y)
\cdot e^{i k_z z}
\right],
\end{equation}
where $k_z = \sqrt{k^2 - k_x^2 - k_y^2}$ accounts for the exact non-paraxial defocus. The factor $1/\sqrt{\cos\theta}$ arises from two contributions: the $1/\cos\theta$ Jacobian from projecting the spherical reference surface onto the planar pupil coordinates and the $\sqrt{\cos\theta}$ amplitude factor required by the Abbe sine condition for aplanatic apodization. The $3\times3$ matrix $\mathbf{M}$ describes the polarization transformation upon refraction by the high-NA lens. For the transverse incident field, the matrix $\mathbf{M}$ can be written as
\begin{equation}
\label{eq:Matrix_M}
\mathbf{M} =
\begin{pmatrix}
1 + (\cos\theta - 1)\cos^2\phi & (\cos\theta - 1)\sin\phi\cos\phi & 0 \\
(\cos\theta - 1)\sin\phi\cos\phi & 1 + (\cos\theta - 1)\sin^2\phi & 0 \\
-\sin\theta\cos\phi & -\sin\theta\sin\phi & 0
\end{pmatrix} .
\end{equation}
In Eq.~\eqref{eq:Matrix_M}, the pupil angles satisfy $\theta\in[0,\theta_{\max}]$ and $\phi\in[0,2\pi)$, with $\sin\theta_{\max}=\mathrm{NA}$, and are related to the spatial frequencies by $k_x=k\sin\theta\cos\phi$ and $k_y=k\sin\theta\sin\phi$. Eq.~\eqref{eq:Formula_Vectorial} is the basis of our vectorial propagation model for the aplanatic microscope-objective geometry considered here \cite{GuAdvancedImaging}.

The effects of polarization and non-paraxial geometry can be isolated by reducing Eq.~\eqref{eq:Formula_Vectorial} into two simplified scalar approximations. First, we keep the same scalar pupil function $E_{\text{in}}(k_x,k_y)=A(k_x,k_y)\exp[i\varphi_{\text{SLM}}(k_x,k_y)]$ but discard the polarization transformation by replacing the vector mapping $\mathbf{M} \cdot\mathbf{E}_{\text{in}}$ with the scalar pupil field $E_{\text{in}}$. This results in the scalar Debye model
\begin{equation}
\label{eq:Formula_ScalarDebye}
    E_{\text{D}}(\mathbf{r}) \propto \mathcal{F} \left[ \frac{1}{\sqrt{\cos\theta}} E_{\text{in}}(k_x, k_y) e^{i k_z z} \right].
\end{equation}
This approach retains the accurate aplanatic amplitude mapping while treating the light field as a scalar wave.

Second, by taking the paraxial limit ($\theta \to 0$), the apodization factor approaches unity ($1/\sqrt{\cos\theta} \approx 1$). If we further evaluate the field strictly at the focal plane ($z=0$), this yields the standard Fraunhofer (paraxial scalar) propagation, used in many traditional holographic algorithms
\begin{equation}
\label{eq:Formula_Fraunhofer}
    E_{\text{F}}(\mathbf{r}) \propto \mathcal{F} \left[ E_{\text{in}}(k_x, k_y) \right].
\end{equation}
Proportionality constants in Eqs.~\eqref{eq:Formula_Vectorial}, \eqref{eq:Formula_ScalarDebye}, and \eqref{eq:Formula_Fraunhofer} are omitted, as the metrics considered are invariant with respect to them. By comparing these three models, we can illustrate the specific limitation of scalar approximations in high-NA holographic setups. Furthermore, implementing these propagation models as FFT operations within a differentiable framework enables computationally efficient gradient-based optimization of the SLM phase.

\paragraph{Hologram optimization} To determine the phase modulation $\varphi(\mathbf{r}_{\text{SLM}})$ required to shape the target intensity $I_{\text{tar}}$, we formulate the problem as a nonlinear optimization task. Traditional iterative projection algorithms, such as Gerchberg-Saxton, require alternating constraints in the SLM and target planes. Extending this projection picture to the RW formalism is not straightforward. First, inverse RW focusing is itself nontrivial, and deterministic inversion formulas require additional assumptions on the specified focal-plane field~\cite{Foreman2008InverseDebyeWolf}. Second, the targets considered here are not prescribed complex vectorial fields, but derived quantities such as total intensity and, later, optical dipole potentials. Replacing a scalar amplitude in the target plane, therefore, does not define a unique update for the vectorial focal field. We avoid this ambiguity by directly minimizing a cost function $\mathcal{L}$ with respect to the phase using gradient-based optimization. We define the objective function as the squared error between the target and simulated intensity, weighted spatially by a mask $\mathcal{M}$. For all intensity optimizations, $\mathcal{M}$ is a circular focal-plane mask centered on the target with a radius of 250 focal-plane pixels ($\simeq 40\,r_{\mathrm{Airy}}$) on the discretized focal-plane grid, chosen to include the full target region while excluding the empty background. To ensure the optimization is invariant to the total source power, we normalize the intensities using the Frobenius norm. The loss function is defined as

\begin{equation}
\label{eq:loss_function}
\mathcal{L}(\varphi)
=
\left\|
\frac{I_{\text{sim}}(\varphi) \odot \mathcal{M}}{\| I_{\text{sim}}(\varphi) \odot \mathcal{M} \|_F}
-
\frac{I_{\text{tar}} \odot \mathcal{M}}{\| I_{\text{tar}} \odot \mathcal{M} \|_F} \right\|_F^2,
\end{equation}
where $\odot$ denotes the element-wise Hadamard product and $\| \cdot \|_F$ is the Frobenius norm. The simulated intensity $I_{\text{sim}}(\varphi)$ is obtained via the differentiable forward models described by Eqs.~\eqref{eq:Formula_Vectorial}, \eqref{eq:Formula_ScalarDebye}, and \eqref{eq:Formula_Fraunhofer}. For scalar forward models, $I_{\text{sim}}=|E_{\text{F}}|^2$ or $|E_{\text{D}}|^2$, whereas in the fully vectorial RW case $I_{\text{sim}}=|\mathbf{E}|^2=|E_x|^2+|E_y|^2+|E_z|^2$. Minimizing Eq.~\eqref{eq:loss_function} requires the phase gradient $\nabla_\varphi\mathcal{L}$. While the analytic gradients of the underlying linear transformations can be derived, manually implementing the complex chain rules across the vectorial propagation is prone to algebraic errors. Instead, we use automatic differentiation in JAX \cite{jax2018github}, which differentiates the implemented discrete model up to floating-point round-off. For the numerical minimization, we use L-BFGS~\cite{LiuNocedal1989LBFGS}, a limited-memory quasi-Newton method that provides deterministic full-batch updates for the large phase vector. Unless stated otherwise, all phase optimizations were run for 1000 L-BFGS iterations from the same initialization protocol, and the reported reconstructions use the final iterate. A compact optimizer-and-normalization sensitivity check is provided in Supplement 1.

\subsection{Numerical implementation and evaluation metrics}
\label{sec:targets}

For the calculations, we restrict the active SLM region to a circular pupil with a radius of 200 pixels, corresponding to approximately $1.2\times 10^5$ phase variables. The pupil is zero-padded in a $2048\times2048$ computational grid, so the FFT-based propagator returns the focal field on the corresponding $2048\times2048$ output grid. All calculations are performed using double-precision complex floating-point arithmetic.
To evaluate the fidelity of the resulting intensities, we define the uniformity
\begin{equation}
    \label{eq:uniformity}
    u = 1 - \frac{\sigma_{\text{SR}}(I_{\text{sim}})}
    {\langle I_{\text{sim}}\rangle_{\text{SR}}}.
\end{equation}
In Eq.~\eqref{eq:uniformity}, both the standard deviation $\sigma_{\text{SR}}$ and the mean $\langle \cdot \rangle_{\text{SR}}$ are computed only over the signal region (SR). For the flat-top target, we also define the mean-normalized intensity
\begin{equation}
    \label{eq:mean_normalized_intensity}
    \tilde{I}_{\text{sim}}(\mathbf{r}) =
    \frac{I_{\text{sim}}(\mathbf{r})}
    {\langle I_{\text{sim}}\rangle_{\text{SR}}},
\end{equation}
and report the flat-top peak-to-valley variation
\begin{equation}
    \label{eq:pv_flat}
    PV_{\text{flat}} =
    \max_{\mathbf{r}\in\text{SR}_{\text{flat}}}
    \tilde{I}_{\text{sim}}(\mathbf{r})
    -
    \min_{\mathbf{r}\in\text{SR}_{\text{flat}}}
    \tilde{I}_{\text{sim}}(\mathbf{r}).
\end{equation}

The signal region is target-dependent. For the flat-top profile, $\text{SR}_{\text{flat}} = \{\mathbf{r}\mid I_{\text{tar}}(\mathbf{r})>0.9999\max(I_{\text{tar}})\}$, which isolates the plateau and excludes the steep target edges. For the tweezer array, the uniformity is computed from fitted peak amplitudes: each spot is fitted with a 2D Gaussian, and the fitted peaks $A_j$ define $u_{\text{tweezer}}=1-\sigma(A_j)/\langle A_j\rangle$.

As a separate tweezer-shape metric, we report the ellipticity $\varepsilon=\sigma_x/\sigma_y$, where $x$ is the incident polarization direction. The ellipticity is computed for each tweezer and averaged over the array.
The two target geometries were selected because they are prevalent in cold-atom experiments \cite{LeeArbitraryTraps2014, Kim2024ProgrammablePotential, endres2016, barredo2016, bluvstein2024} and probe contrasting spatial frequency regimes (near-diffraction-limited spots versus continuous flat-tops). To ensure the targets respect the physical diffraction limits, the spatial frequency content of the target must be bounded. For the extended flat-top profile, we convolve a binary square flat-top profile with the RW total-intensity PSF computed at $\mathrm{NA}=0.9$ for $x$-polarized light. The resulting fixed numerical target is used for all propagation models and NA values. The underlying square has side length $w_x=w_y = 40 \cdot r_{\text{Airy}}$, where $r_{\text{Airy}} = 0.61 \lambda / \mathrm{NA}$ is used as the diffraction-length unit at a design wavelength of $\lambda = \SI{532}{\nano\meter}$. Since the focal-plane pixel size scales with the diffraction length, the target dimensions are fixed in units of $r_{\text{Airy}}$ while their physical size changes with NA.

The tweezer array follows a $10 \times 10$ geometry with a lattice spacing of $d = 4.0 \cdot r_{\text{Airy}}$. For the individual tweezer targets, we use symmetric 2D Gaussian profiles with $\sigma_{\mathrm{tar}}=1.05\,\sigma_{\mathrm{PSF,maj}}$, where $\sigma_{\mathrm{PSF,maj}}$ is the larger fitted standard deviation of the same $\mathrm{NA}=0.9$ vectorial PSF. This avoids a scalar Airy-disk target that can be narrower than the physically achievable vectorial PSF along its major axis, because polarization mixing at high NA elongates the focal spot along the incident polarization axis \cite{DornFocalSpot2003}. The same dimensionless Gaussian target array is then used for all NA values. Supplement 1 examines the sensitivity to this Gaussian-width margin and separately validates the RW implementation through low-NA scalar-limit, analytical component-energy, dense Fourier-sum, and zero-padding checks.

\section{Vectorial hologram generation}

To assess the vectorial optimization framework, we benchmark its performance against standard scalar approaches using the targets discussed in Section~\ref{sec:targets}. Each phase is first optimized self-consistently with the forward model indicated in the figures. Unless stated otherwise, scalar-model results in this section refer to those optimized scalar phases after reevaluation with the RW model. Thus, the scalar errors reported below are model-mismatch errors under RW evaluation, not evidence that scalar optimization fails under its own forward model.

\subsection{Vectorial hologram fidelity}

\begin{figure}[htbp]
    \centering
    \includegraphics[width=1\columnwidth]{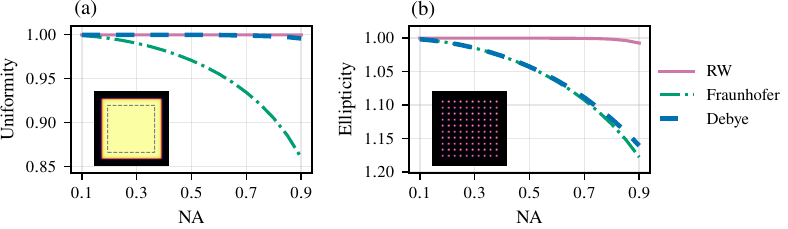}
    \caption{Systematic NA dependence of scalar and vectorial CGH fidelity under RW evaluation. For each numerical aperture, the SLM phase was optimized with the indicated forward model and the reported metric was computed after evaluating the final phase with RW propagation. (a) Uniformity $u$ of the extended flat-top target versus NA. (b) Mean ellipticity $\varepsilon$ of the tweezer-array target versus NA, with $\varepsilon=1$ corresponding to circular traps. The insets show the respective targets, and the dashed gray line in (a) indicates the signal region for the uniformity calculations.}
    \label{fig:Fig2}
\end{figure}

Fig.~\ref{fig:Fig2} summarizes the NA dependence of two representative benchmark metrics for phases optimized with Fraunhofer, Debye, and full RW propagation.

\begin{figure}[htbp]
    \centering
    \includegraphics[width=1\columnwidth]{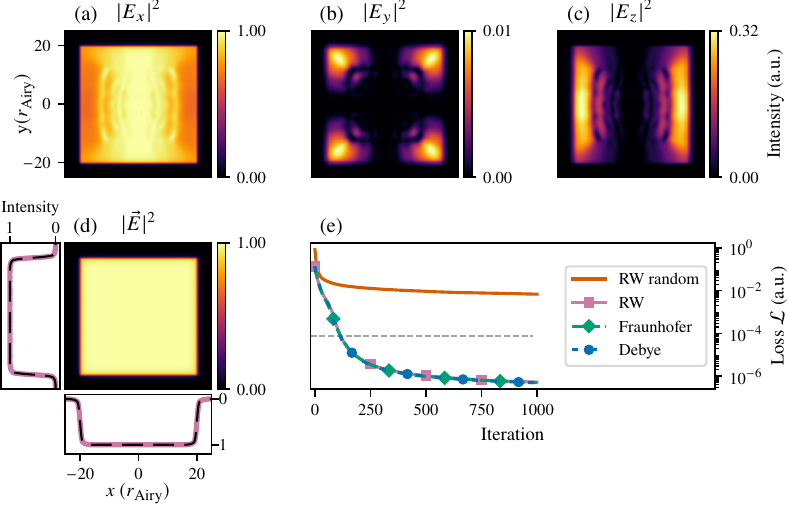}
    \caption{Fully vectorial optimization of an extended flat-top profile at $\mathrm{NA}=0.7$. All intensities are normalized to the mean total intensity in the flat-top signal region, and the component panels use separate color ranges. Panels (a)-(c) show the component intensities $|E_x|^2$, $|E_y|^2$, and $|E_z|^2$. (d) The resulting total intensity landscape ($|\mathbf{E}|^2$). Horizontal and vertical cross-sections (bottom and left margins) compare the optimized solution (solid curve) with the ideal target profile (dashed curve), demonstrating excellent agreement. (e) Self-consistent optimization loss over $\mathcal{M}$ for each run, evaluated with the same forward model used during that run. This panel compares convergence behavior rather than RW-evaluated physical fidelity. The horizontal dotted line marks the RW self-consistent loss at the first iteration where the tracked flat-top uniformity exceeded \SI{99.5}{\percent}.}
    \label{fig:Fig3}
\end{figure}

The trends in Fig.~\ref{fig:Fig2} can be understood by separating scalar high-NA pupil weighting from genuine vectorial polarization mixing. At $\mathrm{NA}=0.7$, the marginal ray angle is $\theta_{\max}\approx\SI{44}{\degree}$, so the aplanatic factor already reaches $1/\sqrt{\cos{\theta_{\max}}}\approx1.18$ at the pupil edge. The RW-optimized flat-top field in Fig.~\ref{fig:Fig3} also contains a sizable longitudinal component, with an integrated fraction $\sum_{\mathbf{r}}|E_z(\mathbf{r})|^2/\sum_{\mathbf{r}}|\mathbf{E}(\mathbf{r})|^2\approx\SI{12.9}{\percent}$ over the field of view. Fraunhofer propagation includes neither the aplanatic weighting nor polarization mixing, whereas the Debye model includes the aplanatic weighting but remains scalar and therefore omits the polarization mixing matrix $\mathbf{M}$.

For the extended flat-top target, Fraunhofer-optimized phase holograms lose uniformity under RW evaluation as NA increases, whereas Debye- and RW-optimized phases remain close to the target. At $\mathrm{NA}=0.7$, standard Fraunhofer optimization followed by RW evaluation gives only $u_{\mathrm F}=\SI{93.43}{\percent}$ and a peak-to-valley variation within $\text{SR}_{\text{flat}}$ of approximately \SI{25.0}{\percent}. In contrast, the Debye model reaches $u_{\text{D}}=\SI{99.90}{\percent}$ and $PV_{\text{flat}}=\SI{0.77}{\percent}$ under RW evaluation. The close Debye-RW agreement identifies the missing aplanatic pupil weighting, rather than the omission of explicit $E_y$ and $E_z$ components, as the dominant source of the Fraunhofer flat-top error. The magnitude of this scalar-model error depends on the pupil mapping and apodization and may differ for other focusing architectures, such as diffractive optical elements.

A spectral view supports this interpretation. For a fixed scalar Debye pupil field and incident $x$ polarization, the RW total-intensity spectrum differs from the Debye spectrum only through a polarization-overlap factor. This factor equals unity at zero intensity spatial frequency, $\mathbf q=0$, and its first nonzero correction is quadratic in $|\mathbf{q}|$; the derivation is provided in Supplement 1. Since the flat-top uniformity metric chosen here mainly probes smooth, low-$|\mathbf{q}|$ variations across the plateau, the Debye and RW total-intensity predictions remain close.

The complementary limitation of the Debye model becomes apparent for structures on the scale of the diffraction limit, such as tweezer arrays, thin structures, or steep walls. For such localized targets, the omission of the polarization mixing matrix $\mathbf{M}$ can lead to geometric distortions that are not captured by scalar high-NA weighting alone. We quantify this effect for the tweezer-array target in Fig.~\ref{fig:Fig2}(b): under RW evaluation, Fraunhofer- and Debye-optimized holograms both develop increasing ellipticity with NA, whereas RW-optimized holograms remain close to circular.
At $\mathrm{NA}=0.7$, a Debye-optimized tweezer hologram under RW evaluation produces elliptical traps with $\varepsilon=1.09$. Incorporating the full RW forward model in the optimization loop suppresses this ellipticity: the SLM phase is updated to actively pre-compensate vectorial polarization mixing, reducing the mean ellipticity to $\varepsilon=1.00$ while maintaining $u_{\text{RW}}=\SI{99.98}{\percent}$.

The calculated $x$, $y$, and $z$ field components for the RW-optimized flat-top hologram at $\mathrm{NA}=0.7$ are shown in Fig.~\ref{fig:Fig3}(a)-(c), with the total intensity $|\mathbf{E}|^2$ shown in Fig.~\ref{fig:Fig3}(d). The horizontal and vertical line cuts in Fig.~\ref{fig:Fig3}(d) show that the total intensity follows the flat-top target despite the sizable longitudinal contribution $|E_z|^2$ shown in panel (c). The convergence curves shown in Fig.~\ref{fig:Fig3}(e) are discussed in the next subsection.

Table~\ref{tab:table_values} gives the full set of RW-evaluated metrics at the largest numerical aperture studied, $\mathrm{NA}=0.9$. For the flat top, the peak-to-valley variation defined in Eq.~\eqref{eq:pv_flat} decreases from \SI{55.49}{\percent} for Fraunhofer optimization to \SI{3.24}{\percent} for Debye and \SI{0.29}{\percent} for RW. For the tweezer array, the fitted peak uniformity alone would not reveal the dominant scalar-model error: Fraunhofer and Debye still give $u_{\text{tweezer}}=\SI{99.35}{\percent}$ and \SI{99.82}{\percent}, respectively, but their RW-evaluated spots are elliptical with $\varepsilon=1.18$ and $\varepsilon=1.16$. The RW-optimized hologram maintains $u_{\text{tweezer}}=\SI{99.98}{\percent}$ while restoring ellipticity to $\varepsilon=1.007$.

\begin{table}[htbp]
    \caption{Quantitative comparison of hologram fidelity metrics at $\mathrm{NA}=0.9$ across different forward propagation models. For each model, the SLM phase was optimized using the indicated forward model (F: Fraunhofer, D: scalar Debye, RW: Richards-Wolf) and subsequently evaluated under the fully vectorial RW model to extract the reported metrics.}
    \label{tab:table_values}

    \centering
    \begin{tabular}{lccc}
    \hline\hline
         & \textbf{F} & \textbf{D} & \textbf{RW} \\ \hline
        $u_{\text{flat}} (\%)$ & 85.99 & 99.56 &  99.97  \\
        $PV_{\text{flat}} (\%)$ & 55.49 & 3.24 &  0.29  \\
       $u_{\text{tweezer}} (\%)$   & 99.35 & 99.82 &  99.98  \\
       $\varepsilon_{\text{tweezer}} $ & 1.18 & 1.16 & 1.007  \\
       \hline\hline
    \end{tabular}

\end{table}

Together, Fig.~\ref{fig:Fig2} and Table~\ref{tab:table_values} show that optimizing the observable $I_{\text{tot}}=|\mathbf{E}|^2$ with the RW forward model suppresses the scalar-model distortions in the total-intensity targets considered here. We optimize $I_{\text{tot}}$ rather than imposing separate targets on $|E_x|^2$, $|E_y|^2$, and $|E_z|^2$, because the focal-field components generated from a phase-only high-NA pupil are not independent design variables. In the RW model, the focal field is constrained by transversality and by the common pupil phase, so arbitrary component-wise intensity targets would generally be overconstrained or physically inconsistent. Optimizing the observable $I_{\text{tot}}$ lets the RW propagator determine the compatible distribution among the three field components.
This motivates using the same computed vector field in objectives beyond the total intensity, as shown in Section~\ref{sec:potentials}.

\subsection{Algorithmic performance and convergence}
\label{sec:convergence}

Having established the physical fidelity of RW-optimized phases, we next examine whether the added vectorial forward model preserves the convergence behavior and computational practicality required for gradient-based CGH. A primary advantage of the scalar Fraunhofer/Debye approach is its reliance on FFT propagation, which scales as $\mathcal{O}(N \log N)$, where $N$ is the number of sampled grid points. Conversely, a direct numerical integration of the exact RW integrals scales as $\mathcal{O}(N^2)$, making it computationally expensive for a large field of view.
In the FFT-based implementation used here, RW forward propagation remains an $\mathcal{O}(N\log N)$ operation. For a fixed input polarization, one vectorial forward pass requires three component-wise FFTs, compared with one FFT for the scalar Fraunhofer and Debye models. Thus, the vectorial model increases the prefactor of the FFT-dominated propagation step but does not change the asymptotic scaling. Wall-clock benchmarks for full optimization steps, including reverse-mode gradient evaluation, are provided in Supplement 1.

Beyond computational cost, the convergence behavior of the vectorial optimization is also relevant. Fig.~\ref{fig:Fig3}(e) shows the self-consistent value of the loss in Eq.~\eqref{eq:loss_function} for each run, evaluated with the same forward model used to compute the corresponding gradients. This comparison is not intended as a physical fidelity metric for the scalar phases, which are assessed separately by RW evaluation in Fig.~\ref{fig:Fig2} and Table~\ref{tab:table_values}. Instead, it tests whether the vectorial RW forward model introduces an obvious convergence penalty in the optimization loop.
All main intensity benchmarks use the same initialization protocol: a low-dimensional Zernike pre-optimization of the defocus term first generates a smooth quadratic starting phase that distributes optical power over the target region, followed by the full high-dimensional L-BFGS phase optimization. With this shared initialization, the RW loss decreases as smoothly as the Fraunhofer and Debye losses, despite the coupling of the three field components. The additional random-start RW curve in Fig.~\ref{fig:Fig3}(e) demonstrates the need for this starting phase optimization. Without this initialization, the final loss is significantly larger, consistent with phase vortices trapping the gradient-based optimizer in local minima.

\section{Optical dipole potentials}
\label{sec:potentials}

The previous sections focused on intensity-based holography, where the objective is to shape the total electric field intensity, $I_{\text{tot}} = |\mathbf{E}|^2$. However, in applications such as atom trapping, the actual optical dipole potential $U(\mathbf{r})$ experienced by the atoms is polarization-dependent. For an atom in a specific state, the potential can be decomposed into scalar, vector, and tensor polarizability parts \cite{PaperKien2012}

\begin{equation}
    U(\mathbf{r}) = -\frac{1}{4} \alpha_s |\mathbf{E}|^2 - \frac{1}{4} \alpha_v \frac{m_J}{2J} \left[ \hat{\mathbf{e}}_B \cdot \operatorname{Im}(\mathbf{E}^* \times \mathbf{E}) \right] - \frac{1}{4} \alpha_t \frac{3 m_J^2 - J(J+1)}{J(2J-1)} \frac{3|\hat{\mathbf{e}}_B \cdot \mathbf{E}|^2 - |\mathbf{E}|^2}{2},
    \label{eq:potential}
\end{equation}
where $\alpha_s$, $\alpha_v$, and $\alpha_t$ are the scalar, vector, and tensor polarizabilities, respectively. $J$ is the total angular momentum quantum number, $m_J$ is its associated magnetic quantum number, and $\hat{\mathbf{e}}_B$ is the unit vector defining the quantization axis. The scalar term is proportional to the total intensity, whereas the vector and tensor terms depend on the local polarization state through $\operatorname{Im}(\mathbf{E}^*\times\mathbf{E})$ and $|\hat{\mathbf{e}}_B\cdot\mathbf{E}|^2$. Matching $|\mathbf{E}|^2$ is therefore not generally equivalent to matching $U(\mathbf{r})$. It is sufficient only when the polarization-dependent terms are negligible or when the focal polarization state is spatially fixed so that the full potential remains proportional to the intensity. In high-NA focusing, longitudinal and cross-polarized field components make this approximation target and atomic-state dependent. Spatial variations in vector and tensor shifts can modify trap depths, trap frequencies, and state-dependent energy shifts.

Because the RW framework computes all focal-field components, we can evaluate $U_{\text{sim}}(\mathbf{E})$ from Eq.~\eqref{eq:potential} and optimize the potential directly. For the attractive Gaussian target, the implementation evaluates the shape loss on the positive trap-depth profiles $-U_{\mathrm{sim}}$ and $-U_{\mathrm{tar}}$; negating both normalized profiles leaves the squared shape mismatch unchanged. We define the normalized potential-shape loss $\mathcal{L}_{\mathrm{shape}}[U_{\text{sim}}(\varphi),U_{\text{tar}}]$ by replacing the target intensity $I_{\text{tar}}$ in Eq.~\eqref{eq:loss_function} with the target optical potential $U_{\text{tar}}$, and the simulated intensity $I_{\text{sim}}$ with the simulated potential $U_{\text{sim}}(\mathbf{E})$. For a stable trap, the designed transverse profile should also be centered on an axial extremum of the potential. Matching only the transverse shape at $z=0$ does not enforce this first-order stationarity condition. We therefore introduce an axial-stationarity regularization term,
\begin{equation}
\label{eq:potential_regularization}
\mathcal{R}_z(\varphi)
=
\frac{1}{N_z}
\sum_{\mathbf{r}_\perp \in M_z}
\left[
\frac{z_0}
{\left\langle
|U_{\mathrm{sim}}(\varphi,\mathbf r_\perp,0)|
\right\rangle_{M_z}}
\left.
\frac{\partial U_{\mathrm{sim}}(\varphi,\mathbf r_\perp,z)}
{\partial z}
\right|_{z=0}
\right]^2,
\qquad
z_0=\frac{\lambda}{2\,\mathrm{NA}^2}.
\end{equation}
Here, $U_{\mathrm{sim}}$ denotes the simulated optical potential before normalization, and $M_z=\{\mathbf{r}_\perp \mid |U_{\text{tar}}(\mathbf{r}_\perp,0)|>0.01\,\max |U_{\text{tar}}|\}$ is the axial-regularization mask, with $N_z$ denoting the number of pixels in this mask. The normalization by the mean absolute potential removes the arbitrary potential-amplitude scale, while the characteristic axial scale $z_0$ makes $\mathcal R_z$ dimensionless. For each optimization, the axial derivative is evaluated using the same differentiable forward model as the corresponding shape loss. Thus, $\mathcal R_z$ penalizes departures from the first-order stationarity condition $\partial_z U_{\mathrm{sim}}=0$ throughout the masked target region.
The potential optimization loss is therefore
\begin{equation}
\label{eq:potential_loss}
\mathcal{L}_{U}(\varphi)
=
\mathcal{L}_{\mathrm{shape}}
[U_{\mathrm{sim}}(\varphi),U_{\mathrm{tar}}]
+
\Lambda_z\,\mathcal{R}_z(\varphi).
\end{equation}
For all three optimizations described below, we set $\Lambda_z=0.4$. We demonstrate this capability for a single-tweezer potential at $\mathrm{NA}=0.7$. The target uses the same Gaussian width parameter as the intensity-tweezer target in Section~\ref{sec:targets}; only the optimized observable changes from intensity to potential. For illustrative purposes and after absorbing the common potential scale, we set $\alpha_s=\alpha_v=\alpha_t=1$, $m_J=1$, $J=1$, and the quantization axis $\hat{\mathbf e}_B=(\hat{\mathbf x}+\hat{\mathbf y})/\sqrt{2}$. The diagonal axis avoids alignment with either transverse coordinate axis. The normalized coefficients define a deliberately polarization-sensitive test case rather than a parameter set for a specific state. The regime is nevertheless physically motivated: in lanthanide atoms, vector and tensor polarizabilities can be sizable, and measured Dy polarizabilities at \SI{532}{\nano\meter} show regimes in which polarization-dependent contributions can be comparable to the scalar contribution after the relevant state and angular factors are included \cite{BlochDy532}.

\begin{figure}
    \centering
    \includegraphics[width=1\linewidth]{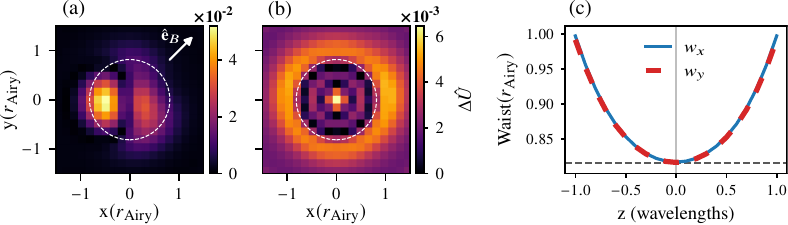}
      \caption{Direct optimization of a polarization-dependent optical dipole potential at $\mathrm{NA}=0.7$. All panels are evaluated with the full Richards-Wolf (RW) field and the optical potential of Eq.~\eqref{eq:potential}. Panels (a) and (b) use the separate color scales shown by their color bars. (a) Normalized absolute residual $\Delta \hat{U}=|\hat U_{\text{sim}}-\hat U_{\text{tar}}|$ for the scalar reference, obtained from a Fraunhofer-optimized hologram and evaluated with the full RW optical potential. The normalized potentials $\hat U$ are mean-subtracted and Frobenius-normalized inside $M_{1/e^2}$ (dashed contour) as defined in Eq.~\eqref{eq:potential_normalization}. The arrow indicates the in-plane quantization-axis direction $\hat{\mathbf e}_B$. (b) Corresponding normalized absolute residual for direct RW-potential optimization, showing strongly reduced potential-shape error under the same full RW evaluation. (c) Through-focus $1/e^2$ potential waists along $x$ and $y$ of the RW-optimized result, extracted from the transverse potential profile in each axial plane. The dashed horizontal line indicates the target waist at $z=0$.}
\label{fig:potential_error}
\end{figure}

We compare three optimization objectives. First, as a scalar baseline, we optimize a Fraunhofer phase using the intensity proxy $U_{\mathrm F}\propto-|E_{\mathrm F}|^2$. For this scalar proxy, $\mathcal R_z$ reduces to the equivalent normalized axial-intensity-gradient penalty. Second, to isolate propagation-model error from objective-function error, we optimize an RW scalar-proxy control using vectorial RW propagation with $\alpha_v=\alpha_t=0$; this control is therefore insensitive to vector and tensor light shifts. Third, we optimize the full polarization-dependent RW potential. All three resulting phases are evaluated with the same full RW potential of Eq.~\eqref{eq:potential}. The resulting transverse residual maps and through-focus waist analysis are shown in Fig.~\ref{fig:potential_error}.

For visualization and quantitative comparison, we first subtract the mean potential inside the $1/e^2$ target-potential mask, and then normalize by the Frobenius norm in the same mask. Thus,
\begin{equation}
\label{eq:potential_normalization}
\hat U =
\frac{U-\langle U\rangle_{M_{1/e^2}}}
{\|(U-\langle U\rangle_{M_{1/e^2}})\odot M_{1/e^2}\|_F},
\quad
M_{1/e^2}=\{\mathbf{r}\mid |U_{\text{tar}}(\mathbf{r})|>e^{-2}\max |U_{\text{tar}}|\}.
\end{equation}
This mean subtraction is used only to report the residual maps and correlation metrics. The optimization loss in Eq.~\eqref{eq:potential_loss} uses the Frobenius-normalized potential shape without subtracting this mean offset. The error maps in Fig.~\ref{fig:potential_error}(a) and (b) show the normalized absolute residual $\Delta \hat{U}(\mathbf{r})=|\hat U_{\text{sim}}(\mathbf{r})-\hat U_{\text{tar}}(\mathbf{r})|$. 

The scalar-reference phase produces clear shape errors and trap asymmetries after the full RW-potential evaluation. These asymmetries are expected for the deliberately polarization-sensitive test case because the final evaluation includes the spatially varying optical spin-density term $\hat{\mathbf{e}}_B\cdot\operatorname{Im}(\mathbf{E}^*\times\mathbf{E})$ and tensor projection $|\hat{\mathbf{e}}_B\cdot\mathbf{E}|^2$, neither of which is constrained by the Fraunhofer intensity-proxy optimization. Direct RW-potential optimization recovers the Gaussian potential shape and suppresses the polarization-induced asymmetry. Inside $M_{1/e^2}$, the mean absolute residual decreases from $2.0\times10^{-2}$ to $1.8\times10^{-3}$, corresponding to a reduction of about a factor of 11, while the Pearson correlation improves from 0.9780 to 0.9998. A Gaussian fit to the RW-optimized potential gives $\varepsilon_U=1.00$, confirming that the recovered potential is circular within the fit precision. This value is used only as a potential-shape diagnostic and is not directly comparable to the intensity-tweezer ellipticities above, because the fit is applied to $U(\mathbf{r})$ rather than to $|\mathbf{E}|^2$. The scalar-only RW control gives a mean normalized residual of $2.1\times10^{-2}$ and a Pearson correlation of $r=0.9768$, close to the Fraunhofer scalar-reference values. Thus, correcting the propagation model while retaining an intensity-proxy objective is insufficient for this polarization-sensitive target; the improvement arises from optimizing the polarization-dependent potential itself.
On the sampled 3D grid, the minimum of the RW-optimized potential occurs at the nominal target position. Any residual displacement of the local trap minimum is therefore below the resolution of the simulated grid. The through-focus waist curves in Fig.~\ref{fig:potential_error}(c) further show that the optimized potential remains nearly symmetric around this plane, consistent with the added axial-stationarity regularization.

\section{Conclusions}

In conclusion, we have demonstrated that integrating an FFT-accelerated Richards-Wolf propagation model with gradient-based optimization enables the generation of high-fidelity computer-generated holograms in the deep non-paraxial regime. Crucially, this approach resolves the target-specific limitations inherent in scalar diffraction models. We identified that Fraunhofer approximations fail for extended flat-top profiles primarily due to the omission of aplanatic apodization. The scalar Debye model largely corrects this pupil-weighting error for extended flat-top profiles, but Debye-optimized holograms exhibit severe degradation for near-diffraction-limited optical structures because they neglect focal polarization mixing. RW optimization suppresses these errors in the total-intensity targets considered here, reaching $u_{\text{flat}}=\SI{99.97}{\percent}$ and $u_{\text{tweezer}}=\SI{99.98}{\percent}$ at $\mathrm{NA}=0.9$ while keeping the tweezer axis ratio close to unity.
Furthermore, we establish that full access to all three electric-field components allows for the direct optimization of polarization-dependent optical dipole potentials. This enables pre-compensation of trap distortions driven by vector and tensor shifts, which conventional intensity-only CGH methods do not capture. This result reframes high-NA CGH as optimization of physically relevant electric-field functionals, not only of scalar intensity.
The present results are computational benchmarks for an ideal phase-only SLM, uniform pupil illumination, and an aplanatic objective; experimental use will require calibration of aberrations, pupil amplitude, SLM pixel response, and polarization optics. A natural next step is to include atom-specific polarizabilities and optimize state-dependent potentials or differential light shifts directly, for example, by minimizing spatial inhomogeneity that contributes to dephasing in neutral-atom traps \cite{Govind2024}.

\begin{backmatter}
\bmsection{Funding}
This work was funded by the European Research Council (ERC) (Grant Agreement No. 101019739) and the Federal Ministry of Research, Technology and Space under the Grant MUNIQC-Atoms (Grant Agreement No. 3N16084).
S.W. acknowledges support from the Center for Integrated Quantum Science and Technology (IQST) and financial support from the German Research Foundation through the Emmy Noether Grant No. WE 7554/1-1, and the Carl-Zeiss-Stiftung Center for Quantum Photonics (QPhoton).

\bmsection{Acknowledgment}
The authors thank Kevin Ng for helpful discussions.

\bmsection{Disclosures} The authors declare no conflicts of interest.

\bmsection{Data availability}
Data underlying the results presented in this paper are not publicly available at this time but may be obtained from the authors upon reasonable request.

\bmsection{Supplemental document}
See Supplement 1 for supporting content.

\end{backmatter}

\bibliography{references}

\clearpage

\setcounter{section}{0}
\setcounter{subsection}{0}
\setcounter{equation}{0}
\setcounter{figure}{0}
\setcounter{table}{0}

\renewcommand{\thesection}{\arabic{section}}
\renewcommand{\thesubsection}{\thesection.\arabic{subsection}}
\renewcommand{\theequation}{S\arabic{equation}}
\renewcommand{\thefigure}{S\arabic{figure}}
\renewcommand{\thetable}{S\arabic{table}}

\renewcommand{\theHsection}{supp.\arabic{section}}
\renewcommand{\theHsubsection}{supp.\arabic{section}.\arabic{subsection}}
\renewcommand{\theHequation}{supp.\arabic{equation}}
\renewcommand{\theHfigure}{supp.\arabic{figure}}
\renewcommand{\theHtable}{supp.\arabic{table}}

\title{High-NA vectorial hologram optimization: supplemental document}

This supplemental document provides numerical validation and additional methodological details for the propagation and optimization results reported in the primary article. It validates the Fraunhofer, scalar Debye, and Richards-Wolf implementations, justifies the Gaussian tweezer target width, derives the intensity-spectrum interpretation, reports timing benchmarks, examines optimizer and normalization sensitivity, and tests a vortex-removal restart for an asymmetric flat-top target.

\section{Numerical Implementation Validation}
\label{sec:validation}

The main results are based on FFT-based implementations of the scalar and vectorial propagation models. This section validates the numerical implementation rather than the physical applicability of the RW model itself, which follows from the standard high-NA focusing formalism used in the main text. All checks in this section are performed using double-precision complex arithmetic on the same phase-only pupil geometry as in the main text.

\begin{figure}[htbp]
\centering
\includegraphics[width=1\linewidth]{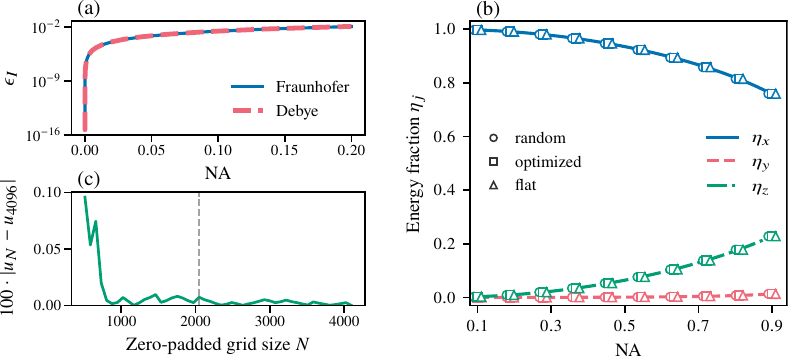}
\caption{Numerical validation of the propagation implementation. (a) Low-NA scalar-limit check. The normalized intensity error $\epsilon_I$ of the Fraunhofer and scalar Debye models is computed relative to the RW result on the full $2048\times2048$ grid. Both scalar models converge to the RW result in the paraxial limit. (b) Integrated RW component-energy fractions for incident $x$-polarized light. Markers show numerical results for a random phase, an optimized flat-top phase, and a flat phase; curves show the analytic phase-independent predictions of Eqs.~\ref{eq:supp_eta_x}-\ref{eq:supp_eta_z}. The random-phase and optimized flat-top markers are shifted horizontally by $\Delta\mathrm{NA}=\pm0.01$ for visibility. (c) Zero-padding convergence of the flat-top uniformity metric. The plotted quantity is the absolute deviation from the $4096\times4096$ reference value in percentage points. The vertical dashed line marks the $2048\times2048$ grid used for the main calculations.}
\label{fig:numerical_check}
\end{figure}

We use four complementary checks. First, we verify that the scalar Fraunhofer, scalar Debye, and vectorial RW propagators approach the same focal-plane intensity in the paraxial limit. Second, we compare the integrated RW component energies with closed-form expressions that are independent of the SLM phase. Third, we verify the FFT-based RW propagator against an explicit dense-matrix evaluation of the same discrete Fourier sum for the sampled RW pupil integrand. Finally, we check that the zero-padded grid size used in the main simulations is sufficient for the reported flat-top uniformity metric. Fig.~\ref{fig:numerical_check} summarizes the low-NA scalar-limit, component-energy, and zero-padding checks; the dense Fourier-sum result is reported in the final paragraph of this section.

\paragraph{Low-NA scalar limit.}
In the limit $\mathrm{NA}\rightarrow 0$, the aplanatic factor approaches unity and the RW polarization matrix approaches the identity on the incident transverse polarization. Therefore, at $z=0$, the Fraunhofer, scalar Debye, and RW models should give the same intensity distribution, up to common normalization constants. We quantify this convergence by comparing sum-normalized intensities over the full $2048\times2048$ computational grid,
\begin{equation}
\label{eq:supp_normalized_intensity_error}
\epsilon_I(A,B)
=
\frac{\|\hat I_A-\hat I_B\|_F}{\|\hat I_A\|_F},
\qquad
\hat I_X(\mathbf r_i)
=
\frac{I_X(\mathbf r_i)}{\sum_i I_X(\mathbf r_i)},
\quad X\in\{A,B\}.
\end{equation}
Here, the first argument denotes the reference intensity. Using the RW intensity as reference, Fig.~\ref{fig:numerical_check}(a) shows $\epsilon_I(\mathrm{RW},\mathrm{F})$ and $\epsilon_I(\mathrm{RW},\mathrm{D})$ for numerical apertures from $\mathrm{NA}=10^{-9}$ to $\mathrm{NA}=0.2$. The same fixed pupil phase is used for all three models. Consistent with the paraxial limit, both scalar models converge toward the RW result as $\mathrm{NA}$ decreases, reaching the numerical round-off level of the calculation at the smallest NA, with $\epsilon_I(\mathrm{RW},\mathrm{F})=4.5\times10^{-16}$ and $\epsilon_I(\mathrm{RW},\mathrm{D})=4.7\times10^{-16}$. The Pearson correlations give the same conclusion, with $1-r_{\mathrm{RW,F}}$ and $1-r_{\mathrm{RW,D}}$ below the displayed numerical precision in the lowest-NA regime. This check confirms that the scalar and RW code paths share a consistent coordinate convention and paraxial limit.

\paragraph{RW component-energy fractions.}
We next validate the vectorial field construction by comparing the integrated component energies with analytic RW predictions. For incident $x$-polarized light, let $M_{jx}$ denote the entries of the first column of the polarization matrix in Eq.~\eqref{eq:Formula_Vectorial} of the primary article. In the focal plane $z=0$, with $\mathbf r=(x,y)$, the $j$-th RW focal-field component can be written as
\begin{equation}
E_j(\mathbf r)
\propto
\mathcal F\left[
E_{\mathrm{in}}(k_x,k_y)
\frac{M_{jx}(\theta,\phi)}{\sqrt{\cos\theta}}
\right],
\qquad
j\in\{x,y,z\},
\end{equation}
where $E_{\mathrm{in}}(k_x,k_y)=A(k_x,k_y)\exp[i\varphi_{\mathrm{SLM}}(k_x,k_y)]$ is the scalar pupil factor. By the Parseval identity,
\begin{equation}
\label{eq:supp_parseval_components}
\int |E_j(\mathbf r)|^2\,d^2r
\propto
\int_{\mathrm{pupil}}
|E_{\mathrm{in}}(k_x,k_y)|^2
\frac{|M_{jx}(\theta,\phi)|^2}{\cos\theta}
\,dk_x\,dk_y .
\end{equation}
For the phase-only pupil with uniform incident amplitude used here, $|E_{\mathrm{in}}|^2$ is independent of the SLM phase. Moreover, with $k_x=k\sin\theta\cos\phi$, $k_y=k\sin\theta\sin\phi$, and $dk_xdk_y=k^2\sin\theta\cos\theta\,d\theta\,d\phi$, the factor $1/\cos\theta$ in~\eqref{eq:supp_parseval_components} cancels the $\cos\theta$ in the planar-pupil Jacobian. The full-field component fractions can therefore be written as
\begin{equation}
\label{eq:supp_eta_angular}
\eta_j
=
\frac{
\int_0^{\theta_{\max}}\int_0^{2\pi}
|M_{jx}(\theta,\phi)|^2\sin\theta\,d\phi\,d\theta
}{
\int_0^{\theta_{\max}}\int_0^{2\pi}
\sum_{\ell=x,y,z}|M_{\ell x}(\theta,\phi)|^2\sin\theta\,d\phi\,d\theta
}.
\end{equation}
This expression is independent of the pupil phase. Evaluating the angular integrals with the same polarization matrix and defining
\begin{equation}
a=\cos\theta_{\max}
=
\sqrt{1-\left(\frac{\mathrm{NA}}{n}\right)^2}
\end{equation}
with $n=1$ for the calculations reported in the main text gives
\begin{align}
\label{eq:supp_eta_x}
\eta_x
&=
\frac{a^2+2a+5}{8},\\
\label{eq:supp_eta_y}
\eta_y
&=
\frac{(1-a)^2}{24},\\
\label{eq:supp_eta_z}
\eta_z
&=
\frac{(1-a)(a+2)}{6}.
\end{align}
These expressions satisfy $\eta_x+\eta_y+\eta_z=1$ exactly and approach $(1,0,0)$ as $\mathrm{NA}\rightarrow0$.

Fig.~\ref{fig:numerical_check}(b) compares Eqs.~\ref{eq:supp_eta_x}-\ref{eq:supp_eta_z} with numerical component fractions obtained by replacing the focal-plane integrals in $\eta_j$ by sums over the entire sampled FFT grid. The comparison is shown for three pupil phases: a random phase, the optimized flat-top phase, and a flat pupil phase. The three numerical data sets overlap within finite-grid sampling error and follow the analytic curves, confirming both the vectorial component construction and the predicted phase independence of the full-field component-energy fractions.

As an additional component-level check, we sum the complex longitudinal field $E_z$ over the complete focal-plane grid. In continuous notation, the transverse integral of $E_z(\mathbf r)$ selects the zero-transverse-frequency value of its pupil integrand: $\int E_z(\mathbf r)\,d^2r\propto E_{\mathrm{in}}(\mathbf 0)M_{zx}(\mathbf 0)$. For incident $x$-polarized light, $M_{zx}=-\sin\theta\cos\phi=-k_x/k$, and therefore $M_{zx}(\mathbf 0)=0$. The full-grid discrete sum is the sampled Fourier-grid analogue of this integral and should therefore vanish independently of the pupil phase. We quantify the remaining numerical residual using the scale-independent ratio $\delta_z=\left|\sum_{\mathbf r}E_z(\mathbf r)\right|/\sum_{\mathbf r}|E_z(\mathbf r)|$. At $\mathrm{NA}=0.9$, $\delta_z$ takes the values $2.2\times10^{-19}$, $2.8\times10^{-19}$, and $1.3\times10^{-18}$ for the random, optimized flat-top, and flat pupil phases, respectively, confirming cancellation of the integrated longitudinal field to numerical precision.

\paragraph{Zero-padding convergence.}
The main simulations use a zero-padded grid of $2048\times2048$ points. To verify that this sampling is sufficient for the reported flat-top uniformity, we repeat the RW evaluation for a fixed optimized flat-top phase while varying the zero-padded linear grid size $N$. For each $N$, the target is resampled to the corresponding focal-plane grid and the flat-top signal region is rebuilt from the resampled target. This avoids comparing different physical regions when the focal-plane pixel size changes with the grid size.

For each grid size, we compute the flat-top uniformity $u_N$ using Eq.~\eqref{eq:Formula_Vectorial} of the primary article. We then compare it with the reference value obtained on a $4096\times4096$ grid,
\begin{equation}
\label{eq:supp_padding_error}
\Delta u_{N} = 100\cdot|u_{N}-u_{4096}|.
\end{equation}
The factor of 100 reports the deviation in percentage points. This is more directly interpretable than a relative error in $u$, because the uniformity is already a normalized quantity close to unity. Fig.~\ref{fig:numerical_check}(c) shows that the uniformity converges rapidly with grid size. At the $2048\times2048$ grid used in the main text, the deviation from the $4096\times4096$ reference is $\Delta u_{2048}=7.2\times 10^{-3}$ percentage points. The small residual oscillations in the sweep are consistent with interpolation and mask discretization effects from rebuilding the finite-pixel signal region at each grid size.

\paragraph{Dense-matrix Fourier-sum cross-check.}
As an implementation check independent of FFT shifts and padding conventions, we evaluate the same sampled RW pupil integral in two numerical ways. The standard implementation computes the component-wise discrete Fourier sums with FFTs, whereas the dense-matrix implementation evaluates the same sums explicitly with exponential matrices. This is not a comparison to a different physical model, but a check that the FFT acceleration reproduces the direct discrete sum of the sampled RW integrand.

Let $\mathbf E_{\mathrm{FFT}}$ and $\mathbf E_{\mathrm{DFT}}$ denote the full vector fields from the two implementations, evaluated on the same focal-plane grid. Before computing the residual, we align the two complex fields by the least-squares global scale factor
\begin{equation}
\alpha
=
\frac{\langle \mathbf E_{\mathrm{DFT}},\mathbf E_{\mathrm{FFT}}\rangle}
{\langle \mathbf E_{\mathrm{DFT}},\mathbf E_{\mathrm{DFT}}\rangle},
\end{equation}
where $\langle X,Y\rangle=\sum_{j,\mathbf r}X_j^*(\mathbf r)Y_j(\mathbf r)$ sums over field components and focal-plane pixels. This removes a possible global amplitude or phase convention. The relative complex-field error is
\begin{equation}
\label{eq:supp_dft_field_error}
\epsilon_E
=
\frac{\|\mathbf E_{\mathrm{FFT}}-\alpha\mathbf E_{\mathrm{DFT}}\|_F}
{\|\alpha\mathbf E_{\mathrm{DFT}}\|_F}.
\end{equation}
For the flat-top phase at $\mathrm{NA}=0.9$ on the full $2048\times2048$ grid, the relative complex vector-field error is $\epsilon_E=2.0\times10^{-14}$. The fitted scale factor is unity to numerical precision, $|\alpha|=1$ and $\arg(\alpha)=-9.6\times10^{-16}$. The corresponding sum-normalized intensity error is $\epsilon_I=3.0\times10^{-15}$. These values are at the expected double-precision round-off level and confirm that the FFT-RW implementation reproduces an explicit dense summation of the same sampled RW Fourier integral. The check therefore validates the FFT acceleration, shift conventions, and component bookkeeping of the discrete RW implementation.

\section{Gaussian Tweezer Target Size}
\label{sec:gauss_sweep}

\begin{figure}[htbp]
\centering
\includegraphics[width=1\linewidth]{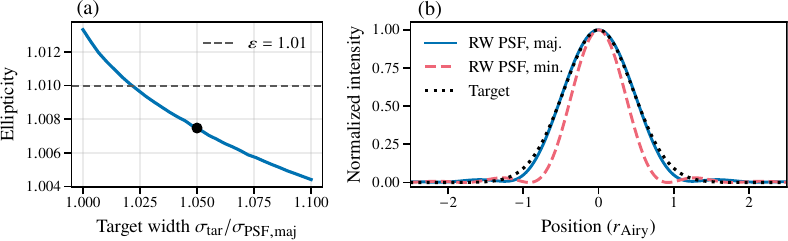}
\caption{Choice of the Gaussian tweezer target width. (a) Mean fitted tweezer ellipticity after RW optimization at $\mathrm{NA}=0.9$ as a function of the symmetric target width normalized to the major-axis width of the vectorial PSF, $\sigma_{\mathrm{tar}}/\sigma_{\mathrm{PSF,maj}}$. The dashed line marks $\varepsilon=1.01$, and the marker indicates the value used in the main text, $\sigma_{\mathrm{tar}}/\sigma_{\mathrm{PSF,maj}}=1.05$. (b) Reference line cuts of the unoptimized RW PSF at $\mathrm{NA}=0.9$ along the incident-polarization direction $x$ and the orthogonal direction $y$, together with the selected symmetric Gaussian target from the marker in panel (a). The larger PSF ellipticity in panel (b), $\varepsilon_{\mathrm{PSF}}\approx1.2$, is the input focal-spot anisotropy that motivates the width sweep; panel (a) shows the residual ellipticity after optimization.}
\label{fig:gauss_sweep}
\end{figure}

We choose the Gaussian tweezer target width used in the main text based on the width sweep shown in Fig.~\ref{fig:gauss_sweep}. For each value of $\sigma_{\mathrm{tar}}/\sigma_{\mathrm{PSF,maj}}$, we optimize the tweezer-array phase with the RW forward model at $\mathrm{NA}=0.9$ and evaluate the mean fitted tweezer ellipticity. The ellipticity deviation falls below \SI{1}{\percent}, $\varepsilon<1.01$, for $\sigma_{\mathrm{tar}}/\sigma_{\mathrm{PSF,maj}} \approx1.02$. We therefore use $\sigma_{\mathrm{tar}}=1.05\, \sigma_{\mathrm{PSF,maj}}$ for the main benchmarks, giving a small margin while keeping the target close to the diffraction-limited vectorial PSF. The line cuts in Fig.~\ref{fig:gauss_sweep}(b) show the selected Gaussian target together with the RW PSF at $\mathrm{NA}=0.9$, illustrating both the elongation of the vectorial PSF along the incident-polarization direction and the modest size of the chosen target broadening.

\section{Intensity Spectrum Analysis}
\label{sec:spectrum}

The main text uses the intensity spectrum to interpret why scalar Debye propagation remains accurate for broad flat-top total-intensity features, while diffraction-scale tweezer shapes are more sensitive to the full vectorial RW model. Here we give the corresponding short derivation. The goal is not to construct a universal scalar transfer function between Debye and RW propagation, but to show how the missing polarization mixing enters the Fourier components of the intensity.

For compact notation in this section, let $\boldsymbol{\kappa}=(k_x,k_y)$ denote the transverse pupil coordinate and let $\mathbf q=(q_x,q_y)$ denote a transverse spatial frequency of the focal-plane intensity. At $z=0$, define the common scalar Debye pupil factor
\begin{equation}
P(\boldsymbol{\kappa})
=
A(\boldsymbol{\kappa})
\exp[i\varphi_{\mathrm{SLM}}(\boldsymbol{\kappa})]
\frac{1}{\sqrt{\cos\theta(\boldsymbol{\kappa})}},
\end{equation}
where the aperture support is included in $A$. With global Fourier constants omitted, the scalar Debye field is
\begin{equation}
E_{\mathrm{D}}(\mathbf r)
=
\int
P(\boldsymbol{\kappa})
\exp(i\boldsymbol{\kappa}\cdot\mathbf r)
\,d^2\kappa .
\end{equation}
For incident $x$-polarized light, the RW field can be written with the same scalar factor multiplied by the first column of the polarization matrix in Eq.~\eqref{eq:Formula_Vectorial} of the primary article,
\begin{equation}
\mathbf E_{\mathrm{RW}}(\mathbf r)
=
\int
\mathbf m(\boldsymbol{\kappa})
P(\boldsymbol{\kappa})
\exp(i\boldsymbol{\kappa}\cdot\mathbf r)
\,d^2\kappa ,
\end{equation}
where $\mathbf m=(M_{xx},M_{yx},M_{zx})^T$. This vector satisfies
\begin{equation}
|\mathbf m(\boldsymbol{\kappa})|^2=1 .
\end{equation}
Thus the RW matrix redistributes a single pupil contribution among $E_x$, $E_y$, and $E_z$, but it does not change the norm of that individual ray.

The difference between the two models appears after forming the intensity. We define the Fourier transform of the focal-plane intensity as
\begin{equation}
\widetilde I(\mathbf q)
=
\int I(\mathbf r)\exp(-i\mathbf q\cdot\mathbf r)\,d^2r .
\end{equation}
Inserting $I_{\mathrm{D}}=E_{\mathrm{D}}E_{\mathrm{D}}^*$ into the definition of $\widetilde I$, inserting the two Debye field integrals, and carrying out the $\mathbf r$ integral gives the scalar Debye intensity spectrum as the pupil autocorrelation
\begin{equation}
\label{eq:supp_debye_spectrum}
\widetilde I_{\mathrm{D}}(\mathbf q)
\propto
\int
P(\boldsymbol{\kappa}_{+})
P^*(\boldsymbol{\kappa}_{-})
\,d^2\bar{\kappa},
\end{equation}
where
\begin{equation}
\boldsymbol{\kappa}_{\pm}
=
\bar{\boldsymbol{\kappa}}\pm\frac{\mathbf q}{2}.
\end{equation}
For the RW total intensity, $I_{\mathrm{RW}}=|E_x|^2+|E_y|^2+|E_z|^2$, the same calculation gives
\begin{equation}
\label{eq:supp_rw_spectrum}
\widetilde I_{\mathrm{RW}}(\mathbf q)
\propto
\int
P(\boldsymbol{\kappa}_{+})
P^*(\boldsymbol{\kappa}_{-})
G(\bar{\boldsymbol{\kappa}},\mathbf q)
\,d^2\bar{\kappa},
\end{equation}
with the polarization-overlap factor
\begin{equation}
\label{eq:supp_overlap_G}
G(\bar{\boldsymbol{\kappa}},\mathbf q)
=
\mathbf m(\boldsymbol{\kappa}_{+})
\cdot
\mathbf m^*(\boldsymbol{\kappa}_{-}) .
\end{equation}
Scalar Debye corresponds to the same expression with $G=1$. At zero intensity spatial frequency,
\begin{equation}
G(\bar{\boldsymbol{\kappa}},0)
=
|\mathbf m(\bar{\boldsymbol{\kappa}})|^2
=
1,
\end{equation}
so Debye and RW have the same integrated total intensity for the same scalar pupil field, up to the common omitted propagation constants.

For small $\mathbf q$, the two pupil points $\boldsymbol{\kappa}_{\pm}$ are close. For the real $x$-polarized case considered here, a Taylor expansion around the midpoint $\bar{\boldsymbol{\kappa}}=(\boldsymbol{\kappa}_{+}+\boldsymbol{\kappa}_{-})/2$ gives
\begin{equation}
\mathbf m\left(\bar{\boldsymbol{\kappa}}\pm\frac{\mathbf q}{2}\right)
=
\mathbf m
\pm\frac{1}{2}D_{\mathbf q}\mathbf m
+\frac{1}{8}D_{\mathbf q}^2\mathbf m
+O(|\mathbf q|^3),
\end{equation}
where $D_{\mathbf q}=\mathbf q\cdot\nabla_{\boldsymbol{\kappa}}$, and all quantities on the right-hand side are evaluated at $\bar{\boldsymbol{\kappa}}$. Twice differentiating $|\mathbf m|^2=1$ along $\mathbf q$ gives $\mathbf m\cdot D_{\mathbf q}^2\mathbf m=-|D_{\mathbf q}\mathbf m|^2$, which reduces the quadratic contribution to the overlap to $-\tfrac{1}{2}|D_{\mathbf q}\mathbf m|^2$. Moreover, $G(\bar{\boldsymbol{\kappa}},\mathbf q)=G(\bar{\boldsymbol{\kappa}},-\mathbf q)$ for real $\mathbf m$, so all odd powers cancel. Hence
\begin{equation}
\label{eq:supp_G_small_q}
G(\bar{\boldsymbol{\kappa}},\mathbf q)
=
1
-
\frac{1}{2}
\left|
\left(\mathbf q\cdot\nabla_{\boldsymbol{\kappa}}\right)
\mathbf m(\bar{\boldsymbol{\kappa}})
\right|^2
+O(|\mathbf q|^4).
\end{equation}
Equivalently,
\begin{equation}
\label{eq:supp_G_quadratic}
G(\bar{\boldsymbol{\kappa}},\mathbf q)
=
1
-
\frac{1}{2}
\left(
g_{xx}q_x^2
+2g_{xy}q_xq_y
+g_{yy}q_y^2
\right)
+O(|\mathbf q|^4),
\end{equation}
with
\begin{equation}
\label{eq:supp_gij_definition}
g_{ij}(\bar{\boldsymbol{\kappa}})
=
\partial_{k_i}\mathbf m(\bar{\boldsymbol{\kappa}})
\cdot
\partial_{k_j}\mathbf m(\bar{\boldsymbol{\kappa}}),
\qquad
i,j\in\{x,y\}.
\end{equation}
The leading RW-Debye correction to the overlap factor is therefore quadratic in the intensity spatial frequency and, in general, anisotropic.

For incident $x$-polarized light, the anisotropy has a simple physical origin. From the same main-text polarization matrix, the longitudinal component of the first column is
\begin{equation}
m_z=-\sin\theta\cos\phi=-\frac{k_x}{k},
\end{equation}
where $k$ is the optical wavenumber in the medium. Hence
\begin{equation}
\partial_{k_x}m_z=-\frac{1}{k},
\qquad
\partial_{k_y}m_z=0 .
\end{equation}
Thus, in the coordinate convention used here, an intensity spatial frequency parallel to the incident polarization, $q_x$, directly probes changes of the longitudinal field component, whereas the orthogonal component, $q_y$, does not do so through $m_z$ to leading order. After averaging over pupil pairs, this makes the low-$q$ correction generally stronger along the incident-polarization direction, consistent with the vectorial ellipticity observed for near-diffraction-limited tweezers.

Eqs.~\ref{eq:supp_debye_spectrum} and \ref{eq:supp_rw_spectrum} are fixed-pupil statements. For any fixed scalar pupil field $P$, the effective ratio
\begin{equation}
H_P(\mathbf q)
=
\frac{\widetilde I_{\mathrm{RW}}(\mathbf q)}
{\widetilde I_{\mathrm{D}}(\mathbf q)}
\end{equation}
is well-defined where $\widetilde I_{\mathrm{D}}(\mathbf q)\neq0$. Near $\mathbf q=0$, it satisfies
\begin{equation}
H_P(\mathbf q)=1+O(|\mathbf q|^2).
\end{equation}
The absence of a linear term follows from the RW polarization overlap and is independent of the SLM phase. The quadratic coefficient of $H_P$, however, is a pupil-pair average weighted by $P(\boldsymbol{\kappa}_{+})P^*(\boldsymbol{\kappa}_{-})$ and is therefore hologram-dependent. Thus, the result explains the low-spatial-frequency agreement for a given Debye pupil field under RW evaluation, but should not be interpreted as a universal transfer function independent of the hologram.

This fixed-pupil result provides the spectral interpretation used in the main text. Agreement at small $|\mathbf q|$ means agreement of the low-pass-filtered total intensity, not pointwise equality of the full focal pattern. In the present flat-top benchmark, the uniformity metric is evaluated on the broad plateau and excludes the steep target boundary; the dominant residual variations are therefore low-spatial-frequency plateau distortions, where $G\approx1$ and Debye and RW total intensities are expected to agree well. Diffraction-scale tweezers, thin features, and steep walls require larger intensity spatial frequencies, corresponding to interference between more widely separated pupil points. For those pupil pairs, the polarization overlap can deviate considerably from one, and scalar Debye can miss the vectorial geometric distortion even when it captures the aplanatic pupil weighting.

Finally, the optimization benchmarks also compare phases optimized separately with different forward models. The fixed-pupil argument above should therefore be read as a mechanism for the observed target dependence, not as a general equivalence theorem for Debye and RW optimization.

\section{Timings}
\label{sec:timings}

The timed results for the flat-top benchmark are shown in Table~\ref{tab:timing}. Propagation, loss, and loss-plus-gradient timings are measured after just-in-time (JIT) compilation and explicit device synchronization. For each entry, we report the mean per-call time over three outer repeats, each containing 100 timed calls. The optimization time is the end-to-end runtime of a complete 1000-iteration Optax L-BFGS optimization from the same fixed Zernike starting phase. All calculations are performed with \texttt{complex128} precision.

The scalar Debye timings are nearly identical to the Fraunhofer timings because both models use one scalar FFT per forward propagation. In contrast, the RW loss-plus-gradient call takes 2.4 times as long as the Fraunhofer call, consistent with the additional vectorial field components and three component-wise FFTs in the forward model. The complete L-BFGS optimization is 1.7 times slower for RW; this smaller factor reflects optimizer overhead and line-search work that are not simply proportional to the isolated loss-plus-gradient timing.

\begin{table}[ht]
\caption{Wall-clock timing benchmarks for the Fraunhofer, scalar Debye, and vectorial Richards-Wolf (RW) propagation models on an NVIDIA RTX 5070 Ti. Prop. denotes one forward propagation call, Loss denotes one scalar loss evaluation, and Loss+grad. denotes one value-and-gradient evaluation of the same loss. These per-call timings are measured after JIT warm-up and synchronization with \texttt{jax.block\_until\_ready}; values are mean $\pm$ sample standard deviation over three outer repeats with 100 timed calls per repeat. Opt. time is the end-to-end wall-clock time for a complete optimization call with 1000 Optax L-BFGS iterations from the same fixed Zernike starting phase, including optimizer bookkeeping and line-search overhead. Rel. grad. denotes the Loss+grad. runtime normalized to the Fraunhofer Loss+grad. runtime, and Rel. opt. denotes the full optimization time normalized to the Fraunhofer full optimization time. The computational grid is $2048\times2048$, the SLM grid is $400\times400$, and all calculations used \texttt{complex128} precision.}
\label{tab:timing}
\centering
\small
\begin{tabular}{lcccccc}
\hline
Model & Prop. (ms) & Loss (ms) & Loss+grad. (ms) & Opt. time (s) & Rel. grad. & Rel. opt. \\
\hline
Fraunhofer & $2.79 \pm 0.01$ & $3.86 \pm 0.01$ & $7.20 \pm 0.02$ & $16.33 \pm 0.07$ & $1.00$ & $1.00$ \\
Debye      & $2.83 \pm 0.03$ & $3.86 \pm 0.01$ & $7.17 \pm 0.07$ & $16.36 \pm 0.10$ & $1.00$ & $1.00$ \\
RW         & $7.04 \pm 0.15$ & $6.98 \pm 0.23$ & $17.23 \pm 0.22$ & $28.06 \pm 0.04$ & $2.39$ & $1.72$ \\
\hline
\end{tabular}

\end{table}

\section{Optimizer and Normalization Sensitivity}
\label{sec:ablation}

The primary article uses L-BFGS with the Frobenius-normalized squared-error objective defined in Eq.~\eqref{eq:loss_function} of the primary article. To check that the reported RW hologram fidelities are not specific to this optimizer or normalization, we repeat the $\mathrm{NA}=0.9$ RW optimizations for the flat-top and tweezer-array targets using either the L-BFGS or Adam optimizer~\cite{KingmaBa2015}, with either Frobenius or sum normalization. With Frobenius normalization, the squared-error objective is, up to a constant factor, equivalent to a cosine-distance objective. We use the squared-error notation for both normalizations because the implemented objective has the same form. For sum normalization, each masked intensity is normalized by its masked sum, and the corresponding squared-error loss is multiplied by the number of masked pixels to keep the numerical gradient scale comparable. This constant factor does not change the minimizer.

For each target, all compared runs use the same pupil, grid size, corresponding masks, double-precision arithmetic, and Zernike/defocus initialization protocol as the main benchmark. The L-BFGS reference uses 1000 iterations. Adam is run for 1500 iterations, giving comparable wall-clock optimization times of approximately 27-29 seconds for all entries in Table~\ref{tab:optimizer_norm_sensitivity}. The Adam learning rate was selected from sweeps for each target and normalization; the values are $0.093$ for both Frobenius-normalized targets, $0.285$ for the sum-normalized flat-top target, and $0.030$ for the sum-normalized tweezer target. All final phases are evaluated with the full RW model and the same physical metrics as in the main text. 

\begin{table}[htbp]
\caption{Optimizer and normalization sensitivity for RW optimization at $\mathrm{NA}=0.9$. All final phases are evaluated with the full RW model. For the flat-top target, the reported shape metric is the peak-to-valley variation $PV_{\mathrm{flat}}$ over the flat-top signal region. For the tweezer array, the reported shape metric is the mean fitted ellipticity $\varepsilon_{\mathrm{tw}}$; peak-to-valley variation is not applied to the tweezer target.}
\label{tab:optimizer_norm_sensitivity}
\centering
\small
\begin{tabular}{lllrrrr}
\hline
Target & Opt. & Norm. & Iter. & $u$ (\%) & $PV_{\mathrm{flat}}$ (\%) & $\varepsilon_{\mathrm{tw}}$ \\
\hline
flat-top & L-BFGS & Frob. & 1000 & 99.97 & 0.29 & -- \\
flat-top & L-BFGS & sum   & 1000 & 99.97 & 0.30 & -- \\
flat-top & Adam   & Frob. & 1500 & 99.87 & 1.80 & -- \\
flat-top & Adam   & sum   & 1500 & 99.68 & 3.45 & -- \\
\hline
tweezer & L-BFGS & Frob. & 1000 & 99.98 & -- & 1.01 \\
tweezer & L-BFGS & sum   & 1000 & 99.97 & -- & 1.01 \\
tweezer & Adam   & Frob. & 1500 & 99.94 & -- & 1.01 \\
tweezer & Adam   & sum   & 1500 & 99.92 & -- & 1.01 \\
\hline
\end{tabular}
\end{table}

The L-BFGS/Frobenius setting used in the main text achieves the best target fidelity among the tested settings. Replacing Frobenius normalization by sum normalization changes the L-BFGS metrics only weakly for both targets, indicating that the main RW results are not tied to the Frobenius normalization used in the primary article. Tuned Adam reaches high-fidelity holograms within a comparable wall-clock budget, but it does not improve the final RW-evaluated target metrics. We therefore use L-BFGS with Frobenius normalization for the main benchmarks.

\section{Suppression of Phase Vortices in High-NA CGH}
\label{sec:vortices}

The main text focuses on connected flat-top profiles and regular tweezer arrays. As a stress test for less symmetric target geometries, we additionally consider an asymmetric split flat-top target obtained by removing the central cross from the square flat-top and omitting one of the four remaining segments before PSF smoothing. For this target, the standard Zernike-initialized RW optimization can converge to local minima containing phase vortices in the dominant focal-field component $E_x$. Vortex-induced stagnation is a known limitation of gradient-based hologram optimization, and related cost-function strategies have previously been used to suppress vortex formation in scalar holographic traps \cite{Harte:14}.

We suppress these defects with an approximate restart procedure. Vortices are detected from the phase $\arg(E_x)$ inside the high-intensity target mask defined by $I_{\mathrm{tar}}>0.8\max(I_{\mathrm{tar}})$. For each detected charge, the corresponding conjugate vortex phase is applied in the focal plane. The corrected phase is then paired with the current total-intensity amplitude $\sqrt{|\mathbf E|^2}$ to form a scalar proxy field, which is backpropagated by an inverse Fourier transform to obtain a new SLM initialization. The restarted L-BFGS run is subsequently performed with the same full RW forward model and intensity loss as in the other optimizations.

\begin{figure}[htbp]
    \centering
    \includegraphics[width=1\linewidth]{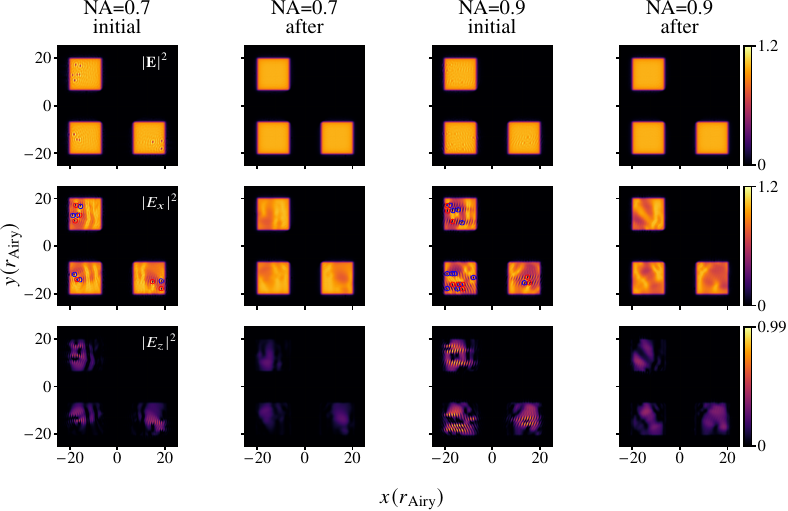}
    \caption{Vortex-removal restart for an asymmetric split flat-top target. Columns show the RW-optimized fields before and after the restart for $\mathrm{NA}=0.7$ and $0.9$. Rows show the total intensity $|\mathbf E|^2$, the dominant transverse component intensity $|E_x|^2$, and the longitudinal component intensity $|E_z|^2$. All intensities are normalized to the mean total intensity in the flat-top plateau of the corresponding reconstruction; color scales are shared within each row. Markers in the $|E_x|^2$ panels indicate phase vortices detected from $\arg(E_x)$ inside the target mask; blue circles and red crosses denote charges $+1$ and $-1$, respectively. The antivortex restart removes the detected $E_x$ vortices and allows the subsequent full RW optimization to recover a uniform total-intensity profile.}
    \label{fig:supp_vortex_removal}
\end{figure}

Fig.~\ref{fig:supp_vortex_removal} shows the result for $\mathrm{NA}=0.7$ and $0.9$. Before restart, the optimized fields contain 12 and 21 detected $E_x$ vortices, respectively. At high NA, the longitudinal component can partially fill the transverse vortex cores, so the total intensity no longer resembles a purely scalar vortex singularity. The plateau metrics nevertheless show that this vortex-containing state remains a poorer local minimum. After the scalar antivortex restart and RW reoptimization, no $E_x$ vortices are detected in the target mask. The flat-top uniformity improves from $\SI{95.17}{\percent}$ to $\SI{99.85}{\percent}$ at $\mathrm{NA}=0.7$, and from $\SI{97.65}{\percent}$ to $\SI{99.85}{\percent}$ at $\mathrm{NA}=0.9$. The corresponding mean-normalized peak-to-valley variations decrease from $\SI{63.66}{\percent}$ to $\SI{1.01}{\percent}$, and from $\SI{22.90}{\percent}$ to $\SI{0.95}{\percent}$, respectively.

\end{document}